\documentclass[12pt]{article}
\usepackage{epsfig} 

\newcommand{\mrm}[1]{\mathrm{#1}}
 
\newcommand{\gtrsim}{\raisebox{-0.8mm}%
{\hspace{1mm}$\stackrel{>}{\sim}$\hspace{1mm}}}
\newcommand{\lessim}{\raisebox{-0.8mm}%
{\hspace{1mm}$\stackrel{<}{\sim}$\hspace{1mm}}}

\newcommand{\MW}{M_{\mrm{W}}}
\newcommand{\GW}{\Gamma_{\mrm{W}}}
\newcommand{\mavg}{\langle m \rangle}
\newcommand{\pavg}{\langle p \rangle}
 
\newcommand{\alphas}{\alpha_{\mrm{s}}}

\renewcommand{\d}{\mrm{d}}
\newcommand{\e}{\mrm{e}}
\newcommand{\f}{\mrm{f}}
\newcommand{\g}{\mrm{g}}

\newcommand{\p}{\mrm{p}}
\newcommand{\q}{\mrm{q}}

\renewcommand{\t}{\mrm{t}}

\newcommand{\C}{\mrm{C}}

\renewcommand{\H}{\mrm{H}}

\newcommand{\W}{\mrm{W}}
\newcommand{\Z}{\mrm{Z}}

\newcommand{\fbar}{\overline{\mrm{f}}}
\newcommand{\pbar}{\overline{\mrm{p}}}

\newcommand{\tbar}{\overline{\mrm{t}}}

\newcommand{\ee}{\e^+\e^-}

\newcommand{\pbarp}{\pbar\p}

\newenvironment{Itemize}{\begin{list}{$\bullet$}%
{\setlength{\topsep}{0.2mm}\setlength{\partopsep}{0.2mm}%
\setlength{\itemsep}{0.2mm}\setlength{\parsep}{0.2mm}}}%
{\end{list}}
\newcounter{enumct}
\newenvironment{Enumerate}{\begin{list}{\arabic{enumct}.}%
{\usecounter{enumct}\setlength{\topsep}{0.2mm}%
\setlength{\partopsep}{0.2mm}\setlength{\itemsep}{0.2mm}%
\setlength{\parsep}{0.2mm}}}{\end{list}}
 
\newlength{\captivewidth}
\newlength{\abstwidth}
\def\thebibliographys#1{\subsection*{References}\list
  {[\arabic{enumi}]}{\settowidth\labelwidth{[#1]}\leftmargin\labelwidth
    \advance\leftmargin\labelsep
    \usecounter{enumi}}
    \def\newblock{\hskip .11em plus .33em minus -.07em}
    \sloppy
    \sfcode`\.=1000\relax}

\begin{document}
 
\sloppy

\pagestyle{empty}

\begin{flushright}
DTP/95/68  \\
LU TP 95--18 \\
hep-ph/9508332 \\
August 1995
\end{flushright}
 
\vspace{\fill}
 
\begin{center}
{\LARGE\bf QED Interconnection Effects on }\\[5mm]
{\LARGE\bf W Momentum Distributions at LEP~2}\\[10mm]
{\Large Valery A. Khoze} \\[2mm]
Department of Physics, University of Durham, \\[1mm]
Durham, DH1 3LE, U.K.\\[3mm]
and \\[3mm]
{\Large Torbj\"orn Sj\"ostrand} \\[2mm]
Department of Theoretical Physics 2, University of Lund, \\[1mm]
S\"olvegatan 14A, S-223 62 Lund, Sweden
\end{center}
 
\vspace{\fill}
 
\begin{center}
{\bf Abstract}\\[2ex]
\vspace{-0.5\baselineskip}
\noindent
\begin{minipage}{\abstwidth}
The process $\ee \to \W^+\W^- \to \f_1 \fbar_2 \, \f_3 \fbar_4$
contains charges in the initial, intermediate and final stages.
This gives a rich selection of possible QED interconnection effects.
Coulomb interaction is the simplest of these, and can thus be used 
to explore consequences. We study a number of experimental 
observables, with emphasis on those related to the $\W$ momentum
distribution. Second-order Coulomb effects are shown to be
practically negligible. The limited LEP~2 statistics will not allow
detailed tests, so any theory uncertainty will be reflected in the
systematic error on the $\W$ mass. Currently the uncertainty from 
this source may be as high as 20~MeV.
\end{minipage}
\end{center}
 
\vspace{\fill}

\clearpage
\pagestyle{plain}
\setcounter{page}{1}

\section{Introduction}

A precision measurement of the $\W$-boson mass $\MW$ is one of the main
objectives of the LEP~2 $\ee$ collider. To exceed the precision of 
$\pbarp$ collider experiments, $\MW$ should be measured to an accuracy 
of 50~MeV or better, see e.g. ref.~\cite{VK1}. The success of
such an ambitious goal relies on an accurate theoretical knowledge of
the dynamics of the production and decay stages in 
$\ee \to \W^+\W^- \to 4$~fermions. Moreover, owing to the large $\W$ 
width $\GW$, these stages are not independent but may be interconnected 
by QED and QCD interference effects, which also must be kept under
precise theoretical control. Interconnection phenomena may efface the
separate identities of the two $\W$ bosons, and the final state may no
longer be considered as a superposition of two separate $\W$ decays.
All such effects related to the radiatively corrected (total and 
differential) $\W^+\W^-$ production cross section should be accurately 
calculated, so that the $\W$ mass can be extracted from data.

For the totally inclusive $\W^+\W^-$ production cross section there is
a general proof \cite{VK2,VK3} (also \cite{VK4}) that the radiative
interconnection effects are suppressed by $O(\alpha\GW/\MW)$ or
$O(\alphas^2\GW/\MW)$. The only exception is the Coulomb interaction
between two slowly moving $\W$'s, which is modified by the instability
effects in the very narrow region just near threshold,
$\sqrt{s} - 2 \MW \lessim \GW$. By contrast, differential distributions 
could be distorted on the level of $O(\alpha)$ or $O(\alphas^2)$.
(Analogous QCD final-state interaction effects appear in the
differential distributions of $\t\tbar$ production processes
\cite{VK3,VK4,VK5,VKR1,VK12}.) These distortions should vanish in the
zero-width limit. In ref. \cite{VK6} we studied the possible impact of
QCD interconnection (colour rearrangement) effects for the
direct reconstruction of the $\W$ mass from the hadronic decay 
products (as favoured by experimenters \cite{VK7}). The limited 
understanding on the hadronization of overlapping hadronic systems 
was shown to imply a systematic error on $\MW$ of 30--40 MeV. Additional 
errors could come from Bose-Einstein effects among the final-state 
hadrons \cite{Leif,Rasmus}.

There are also purely QED final-state interactions, however, that can
produce non-factorizable radiative corrections to the Born cross section
of $\ee \to \W^+\W^- \to 4$~fermions \cite{VKR,VK3}. Specifically, there
is a class of virtual interference effects corresponding to the
so-called charged-particle poles \cite{VK3,VK4,VK8}, which induce an
explicit dependence of the differential cross section on the $\W$-boson 
virtualities. This is related to the fact that the dominant contribution
to the non-factorizable radiative interferences comes from sufficiently
soft photons with $k_{\gamma} \lessim \GW$. 
The phenomenon could be exemplified by the Coulomb interaction 
between two unstable $\W$ bosons \cite{VK2,VK9} (also \cite{VK10}).
(To the best of our knowledge, the necessity to take into account
the effects of Coulomb final-state interactions on the $\W$ momentum
distribution was first pointed out in ref. \cite{VK6}.) 
Coulomb terms could be uniquely separated from the other electroweak
corrections in the threshold region, where they provide the dominant
source of the off-shellness effects. 

Other radiative mechanisms may become essential at larger $\W$-boson
velocities $\beta$, for example, those caused by the
intermediate--final or final--final interferences \cite{VK3,VK8}. 
(We recall, however, that the Coulomb correction is not uniquely
defined and gauge independent in the relativistic region.)
In the ultrarelativistic region it is expected \cite{VK8} that the 
dependence on the $\W$ virtuality disappears in the full expression 
for radiative corrections. This phenomenon is deeply rooted in the 
conservation of a charged current.

In the threshold region the Born cross section depends strongly on
the momentum $p$ of the unstable particle. This leads to a competition
between the Breit-Wigner factors and the rest of the cross section 
($\propto \beta$) \cite{VK6}, which induces a significant 
change between the nominal and the actual average mass. In the energy 
range 
\begin{equation}
\MW \gg E \geq \GW ~,
\label{energyreg}
\end{equation}
where $E = \sqrt{s} - 2\MW$ is the nonrelativistic energy of the two
$\W$'s, the mass shift towards lower values is of 
$O(\GW^2/E)$. This shift changes sign at around 190~GeV because of the
increasing r\^ole of the negative higher-order terms ($\propto \beta^2$)
in the Born cross-section expression \cite{VK11}.

Our interest is in the dynamical QED effects, on top of the purely
kinematical distortions. The Coulomb final-state interaction between
unstable $\W$'s could, in principle, induce a systematic shift in the 
reconstructed
$\MW$ of $O(\alpha \pi \GW)$ \cite{VK9,VK11} that is of the same magnitude 
as the aimed-for precision of LEP~2 measurements. Therefore the shifts in 
the $\W$ momentum distribution is the result of a complicated interplay
between phase-space and virtuality-induced effects appearing in the
radiatively corrected cross section for $\ee \to \W^+\W^- \to 4$~fermions. 
In the main region of LEP~2 running, $\sqrt{s} \simeq 175$~GeV, the two 
effects above are of the same order. To obtain reliable numerical
results one also needs to analyze carefully the r\^ole of other 
final-state interaction mechanisms (for such an attempt see ref. 
\cite{VK8}).

In this paper we analyze the impact of QED interconnection,
exemplified by the Coulomb effects on the $\W$ momentum distribution in
the threshold region. This region is relevant for LEP~2 physics, and 
it is just here that off-shell and finite-width effects are most important
\cite{VK2,VK3,VK9,VK12prm}. It may be seen as a continuation of the
studies presented for mass distributions in ref.~\cite{VK11}.

Particular emphasis is put on the momentum distribution of the $\W$'s.
This is especially relevant since the Coulomb effects depend primarily
on the $\W$ momentum. However, there is also an experimental reason for 
such a choice. In the best of possible worlds, the final state could be 
subdivided into two sets of particles, one stemming from the $\W^+$
decay and the other from the $\W^-$. The invariant masses $m_1$ and 
$m_2$ could then be calculated and directly compared with theory.
In addition to the assignment problems already mentioned, such an 
approach suffers from the limited energy resolution of detectors. 
Jet directions, on the other hand, are much better reconstructed.
If energy and momentum conservation relations can be used, i.e.
if jet energies are suitably rescaled to the known $\ee$ beam energy,
the $\W$ mass is better constrained \cite{VK7}. Optimal use is made of 
the experimental information if the $\W^+$ and $\W^-$ of each event 
are assumed to have the same mass, i.e. if one single average mass 
$\overline{m}$ is extracted for each event, rather than $m_1$ and
$m_2$ separately. Such a mass measurement obviously is equivalent
to a momentum measurement, according to the relation 
$\sqrt{s} = 2 \sqrt{\overline{m}^2 + p^2}$. However, the
momentum is the more direct observable, since it is related to the
acollinearity of the jet pairs; we recall that the limit $p=0$
corresponds to each hadronic $\W$ decay giving a pair of back-to-back 
jets. A measurement along these lines is therefore excluded only when 
both $\W$'s decay leptonically.   

The results presented here are aimed at the rather modest task of 
evaluating the size of the QED interconnection effects. They should 
only be considered as a qualitiative guide rather than a complete 
numerical prediction. In what we follows, incoming electrons and 
positrons are assumed unpolarized, and initial-state radiation (ISR) 
effects are neglected. We also assume that all particles can be 
perfectly well measured and neglect the QCD interconnection effects.

In section~2 some of the basic Coulomb formulae are reviewed, with
emphasis on the physical origin and consequences. Section~3 contains
a numerical study of several observables relevant at LEP~2,
with conclusions on the practical importance of Coulomb effects,
on the impact of theoretical uncertainties, and on the choice
of experimental observables. Finally, section~4 contains a 
summary and an outlook.      

\section{Qualitative discussion}

For illustrative purposes we present here the results for momentum
distributions in the non-relativistic approximation. The numerical
calculations in the next  section, however, are based on the full 
relativistic formulae of ref.~\cite{VK9}.

As shown in ref.~\cite{VK12} (see also ref.~\cite{VK9}) the 
non-relativistic Green's function describing the interaction of the $\W$'s
depends only on one off-shell variable, namely the $\W$ momentum $p$
($p = |\mathbf{p}^+| = |\mathbf{p}^-|$ in the rest frame of the pair).
The integration over the $\W^{\pm}$ squared masses $s_1$ and $s_2$
reduces to an integration over $p^2$ with
\begin{equation}
p^2 = \frac{(s - s_1 - s_2)^2 - 4 s_1 s_2}{4s}
    \simeq (\sqrt{s} - \sqrt{s_1} - \sqrt{s_2}) \MW ~.
\end{equation}
In the dominant $p$ region the differential distribution for
$\ee \to \W^+\W^- \to \f_1\fbar_2\,\f_3\fbar_4$ can be written in 
terms of the rapidly changing variable
\begin{equation}
x = \frac{\MW E - p^2}{\MW \GW}
\end{equation}
as
\begin{equation}
\frac{1}{B(\W^+ \to \f_1\fbar_2) \, B(\W^- \to \f_3\fbar_4)} \, 
\frac{\d \sigma}{\d x} = \frac{\sigma_0}{\pi} \, \frac{1}{1+x^2} \,
\left[ \left( 1 + \frac{\alpha}{\pi} \, \delta_{\H} \right) +
\frac{\alpha}{\beta} \, \overline{\delta}_{\C} \right] ~,
\label{sigmarad}
\end{equation}
with $\beta = 2 p /\sqrt{s}$. Here $\delta_{\H}$ is the so-called hard 
correction and $\alpha\overline{\delta}_{\C}/\beta$ is the first-order 
Coulomb term
\cite{VK2,VK9},
\begin{equation}
\overline{\delta}_{\C} = \frac{\pi}{2} - 
\arctan \frac{|\kappa|^2 - p^2}{2 p p_1} ~,
\label{deltaC}
\end{equation}
with 
\begin{equation}
\kappa = \sqrt{-\MW (E + i\GW)} \equiv p_1 - i p_2 ~.
\label{kappa}
\end{equation}
For a complete calculation, one must add to eq.~(\ref{sigmarad}) the
contributions from other final-state interconnection terms.

For the energy region (\ref{energyreg}) one finds
\begin{equation}
\overline{\delta}_{\C} = \frac{\pi}{2} - \arctan x~.
\label{arctanx}
\end{equation}
In this region $p_1$ and $p_2$ have very transparent interpretatations:
\begin{equation}
R_{\tau} \sim \frac{1}{p_1} \sim \frac{\beta_0}{\GW}
\end{equation}
(with $\beta_0 = p_0/\MW \simeq \sqrt{E/\MW}$) is the typical spatial
separation between the diverging quasi-stable $\W$'s, while
\begin{equation}
R_{\C} \sim \frac{1}{p_2} \sim \frac{1}{p_0} \sim \frac{1}{|\kappa|}
\end{equation}
is the characteristic distance of the Coulomb interaction between 
on-mass-shell $\W$ bosons. Thus the dominant contribution to
$\overline{\delta}_{\C}$ comes from values $r \lessim 1/p_0$, where
$p_0 = \sqrt{s/4 - \MW^2}$ and
$r$ is the relative distance between the $\W$'s, as discussed in
detail in ref. \cite{VK9}. $\overline{\delta}_{\C}$ depends
crucially on the virtuality, and is controlled by the phase shift 
between the spatial oscillations corresponding to the actual momentum
$p$ and the characteristic momentum $p_2$. Let us briefly discuss the
main messages of an analysis of this momentum distribution.
\begin{Enumerate}
\item For stable $\W$'s the on-the-mass-shell Coulomb term would always
induce a shift of the distribution towards lower momentum values.
In this limit, the exact Coulomb result \cite{VK13} is actually
twice the first-order contribution ($\pi/2$ in eq.~(\ref{deltaC})).
\item For unstable $\W$'s $\overline{\delta}_{\C}$ is always below
the on-the-mass-shell value of $\pi/2$ for $p < |\kappa|$ and exceeds
it for $p > |\kappa|$. The transition occurs dominantly when
\begin{equation}
| p - p_2 | \lessim p_1~.
\end{equation} 
For region~(\ref{energyreg}) this corresponds to $|x| \lessim 1$.
When $p < p_0$ the effects of destructive interference accumulate
in the course of integration over distances $r$ up to \cite{VK9}
\begin{equation}
r \lessim \frac{1}{p - p_0} < \frac{\beta_0}{\GW}
\end{equation}
and, as a result, $\overline{\delta}_{\C}$ gets screened. 
Correspondingly, when $p > p_0$, there is a constructive interference
in the same range of $r$ values, leading to an increase of 
$\overline{\delta}_{\C}$.
\item In the formal limit $\MW \gg |\kappa| \gg p$ one has
\begin{equation}
C = \frac{\alpha}{\beta} \, \overline{\delta}_{\C} \simeq
\alpha \, \frac{2 p_1}{\sqrt{E^2 + \GW^2}} ~.
\end{equation}
Contrary to the stable case, the Coulomb correction $C$ does not blow up
at $\beta \to 0$ but reaches its maximal value of
$\alpha \sqrt{2\MW/\GW}$ at $E=0$. Below threshold, when $|E| \gtrsim \GW$,
\begin{equation}
C \simeq 2 \alpha \, \sqrt{\frac{\MW}{|E|}}~.
\end{equation}
In region~(\ref{energyreg}), relevant for LEP~2, Coulomb 
corrections are suppressed,
\begin{equation}
C \simeq \frac{\alpha}{\beta_0} \, \frac{\GW}{E} ~.
\end{equation}
The screening of the $1/p$ singularity for $p \ll p_0$ also takes place
for higher-order Coulomb effects. This can be explicitly seen from the
general formulae presented in ref.~\cite{VK9}. The higher-order
Coulomb terms could in principle be more important numerically for 
differential distributions than for the total cross section;
see however the discussion in the next section.
\item $\overline{\delta}_{\C} \to \pi$, i.e. twice the leading-order 
on-the-mass-shell result, when $\MW \gg p \gg |\kappa|$.
However, when $E \gg \GW$ and $\MW \gg p \gg p_0$ the Coulomb correction
$\alpha\pi/p$ formally is much smaller than the on-the-mass-shell value
$\alpha\pi/(2p_0)$. In region~(\ref{energyreg}) the arctan modification
of $\overline{\delta}_{\C}$ in eq.~(\ref{arctanx}) is an odd function
of $x$, so after integration over $p$ the total cross section agrees
with the stable-$\W$ result.
\end{Enumerate}

The instability effects induced by Coulomb interactions are expected to
be reduced at relativistic energies because of the contributions of
other final-state interaction mechanisms. We will return to this issue
below.

\section{Numerical results}

In the following calculations, we assume a W mass $\MW = 80.41$~GeV, 
based on the recent CDF number \cite{CDF}, so as to simplify 
comparisons with the results of ref.~\cite{VK11}. The width
is then $\GW = 2.1$~GeV. Results are presented in the energy
range 150--200~GeV, and for six discrete 
energies: 10 GeV and $\GW$ below threshold, at threshold, $\GW$
above threshold, at 175 GeV (typical LEP~2 running energy) and
30 GeV above threshold. Initial-state photon radiation is neglected, 
so as not to confuse the issue --- in an idealized world this would be 
obtained by the process $\gamma\gamma \to \W^+\W^-$. Other 
interfering graphs, such as $\ee \to \Z^0\Z^0 
\to 4$~fermions, are not considered. Only trivial loop corrections 
are included, such as the running of $\alpha$ and $\alphas$. The 
lowest-order cross section $\sigma_0$ is therefore given by the 
Muta~et~al. expression \cite{MUTA}.

Several comparisons will be presented between alternative descriptions
of Coulomb effects. This will help us assess the impact of QED
interconnection, and to estimate uncertainties. 
\begin{Itemize}
\item The simplest possible alternative is to have no Coulomb 
corrections at all, henceforth designated ``no Coulomb''. Many studies 
in the past have been based on this approach, i.e. using $\sigma_0$
unmodified.
\item In the first-order ``stable Coulomb'' description, the two $\W$ 
bosons are 
considered as stable particles. Then $\sigma_0$ is modified by 
a simple factor $1 + \alpha \overline{\delta}_{\C} / \beta = 
1 + \pi\alpha/2\beta$.
\item The ``unstable Coulomb'' approach is a realistic 
standard for the effects of unstable $\W$'s. Again the multiplicative
factor is $1 + \alpha \overline{\delta}_{\C} / \beta$, but now 
$\overline{\delta}_{\C}$ is given by eq.~(\ref{deltaC}). 
\item Recently simple formulae were presented for the calculation of 
the Coulomb effects for unstable W's to second order \cite{VK14}, 
``second-order Coulomb''. The cross section is now given by 
$\sigma_0$ multiplied by a factor
$|f(\mathbf{p},E)|^2$, with
\begin{equation}
f(\mathbf{p},E) = 1 + \frac{\alpha\sqrt{s}}{4 i p} \, \ln D +
\frac {\alpha^2 s}{16 i p \kappa} \int_0^1 \frac{\d z}{z} \,
\ln \frac{1 + z D}{1 + z / D} ~,
\end{equation} 
where $D = (\kappa + i p)/(\kappa - i p)$ and $\kappa$ is defined in
eq.~(\ref{kappa}).
\end{Itemize}

Fig.~1 compares the total $\ee \to \W^+\W^-$ cross section in the four
scenarios above as a function of c.m. energy. The differences are more 
readily visible in Fig.~2, where the cross sections have been normalized
to the no-Coulomb one. The corrections reach a maximum at around 
threshold, of about 6\%, and thereafter drop, asymptotically like
$1/\beta_0$. They also drop below threshold. This latter behaviour 
may be understood from the shape of the differential momentum 
distribution $\d\sigma / \d p$ in the four scenarios above, Fig.~3. 
Note the very broad momentum spectra at energies below the threshold, 
where (at least) one of the $\W$'s is pushed significantly off the mass 
shell, and thus has a wide mass distribution. This depletion of
the low-$\beta$ region thus reduces the Coulomb corrections below 
threshold. Above threshold the spectrum is peaked close to the 
on-mass-shell value $p_0$ with a width decreasing like 
$\delta p \sim \MW \GW / p_0$. 

As above, differences between the alternative Coulomb scenarios are 
more easily seen it the ratio of $\d\sigma / \d p$ distributions with
and without Coulomb effects included, Fig.~4. In particular note
the characteristic destructive interference for $p < p_0$ in the
unstable Coulomb case. 

Even with this expanded scale, the difference between first- and  
second-order unstable Coulomb is very small. To further quantify
effects, Fig.~5 gives the ratio of the second-order contribution
to the first-order one. At energies below and around threshold,
the effects are positive but only a few per cent, and vary slowly 
with $p$. Above threshold, the difference between low and high $p$
is noticeable on this scale. For $p > p_0$ the effects are close to 
vanishing, $< 1$\%, but become negative of $O(10\%)$ for
for $p < p_0$. In the limit $p \ll p_0$ a simplified analytical
expression is
\begin{equation}
|f(\mathbf{p},E)|^2 = 1 + \frac{2 \alpha \MW p_1}{|\kappa|^2}
\left\{ 1 + \frac{\alpha \MW}{2 p_1} \left[ 1 + 2 \ln 2
\frac{p_1^2 - p_2^2}{|\kappa|^2} \right] \right\} ~.
\label{limitsmallp} 
\end{equation}
This gives a ratio indicated by a dashed line in Fig.~5.
The sign of the second-order effect is such as to further reduce
the next-to-vanishing first-order Coulomb corrections for $p < p_0$. 
We must therefore conclude that it will be impossible to observe any 
second-order Coulomb effects in the differential distributions
at LEP~2. This alternative will largely 
be dropped from the continued discussion, although it is included in 
the figures.

As noted earlier, the contributions of other intermediate--final and
final--final interaction mechanisms have been neglected in the above 
scenarios, either the calculation is first- or second-order. 
Driven by physical intuition we might try to estimate the possible 
impact of such screening corrections by using a simplified formulae
containing an extra factor $(1-\beta)^2$ in front of the arctan term 
in eq.~(\ref{deltaC}) of the first-order unstable
Coulomb formulae, ``dampened Coulomb''. This is 
in agreement with the results of ref.~\cite{VK8}, where an attempt 
is made to perform the quantitative studies of the instability effects 
in the relativistic region. However, it should be noted that the 
asymptotic behaviour does not constrain the form for LEP~2 energies, 
where $\beta \simeq 1/2$, and thus the numerical difference between 
two asymptotically equivalent forms such as $(1-\beta)^2$ and 
$(1-\beta^2)^2$ (or $(1-\beta)^2$ and $(1-\beta_0)^2$) is very large.
The curves for this alternative at 175 and 190 GeV in Fig.~4 should
therefore be taken as a trial only. 

The ansatz above gives a Coulomb effect that is essentially constant 
as a function of $p$. Around $p_0$ the sharp increase noted in
the unstable Coulomb case is considerably dampened, since 
$(1-\beta)^2 \sim (1-0.5)^2 = 0.25$. The visual impression is of an 
even stronger dampening than just a factor 4, since the general 
$1/\beta$ fall-off comes in addition and is common. Formally,
the dampened Coulomb option can be seen as a mixture of $(1-\beta)^2$ 
parts of the unstable Coulomb and $1-(1-\beta)^2$ of the stable Coulomb,
which offers an alternative approach to understanding the r\^ole of the
$1/\beta$. Since the fraction of the stable Coulomb vanishes like
$2\beta - \beta^2$ for $\beta \to 0$, but the Coulomb factor increases
like $1/\beta$, the flat behaviour for small $p$ is to be expected.
This should not be given too much emphasis, however, since we already 
noted that the whole approach is unreliable for the LEP~2 energy region 
and in particular for momenta that deviate
too much from $p_0$. However, in summary, it must be concluded that
the dampened Coulomb scenario looks rather much like the no-Coulomb
ones, except that the cross section is larger. So long as we are
only concerned with distributions normalized per event, no separate 
figures need therefore be given for this alternative. This conclusion
is not valid around or below threshold, but numerically is a sensible
approximation from 170~GeV onwards.

We now proceed to quantify how much event properties are changed by
Coulomb effects. Fig.~6 shows the average $\W$ momentum as a function
of c.m.~energy. At large energies this is increasing like 
$\pavg \approx \sqrt{s/4 - \MW^2}$, but around threshold one or both 
$\W$'s are pushed off mass shell and so $\pavg$ stabilizes at a 
level around 20~GeV. The addition of Coulomb effects gives a shift 
of around 120~MeV in the threshold region, Fig.~7. The effect is here 
negative, i.e. the Coulomb factors enhance the low-momentum tail of 
the spectrum (cf. Fig.~4), giving a reduced $\pavg$. From 164~GeV 
onwards the destructive interference at low momenta leads to a net 
$\pavg$ shift upwards in the unstable Coulomb case. Lower momenta
are always favoured for the (unrealistic) stable Coulomb alternative,
so the asymptotic convergence towards the no-Coulomb baseline
is from opposite directions for a description in terms of stable
or unstable $\W$ bosons. 

Fig.~8 shows the shift between the average $\W$ mass of events and 
the nominal $\MW$ mass parameter (cf.~ref.~\cite{VK11}). Around and 
below threshold this is 
a negative number, while it becomes positive at around 190~GeV.
This is explained as a combination of phase space and cross section 
effects, as discussed in the introduction. Coulomb effects on the
average $\W$ mass $\mavg$ are opposite in sign and about half as large 
as those on the average $\W$ momentum $\pavg$, Fig.~9. This is easily 
understood from the relation \mbox{$p^2 + m^2 \approx s/4$} 
$\Longrightarrow$ \mbox{$\delta m \approx - p \delta p /m$}, 
where \mbox{$1/4 \lessim \pavg/m \lessim 3/4$} for the energy range 
considered here. The mass difference betweem first- and second-order
Coulomb effects is seen to be below 3~MeV everywhere, i.e. completely
negligible.

Since each event contains two $\W$'s, it is possible to study
separately the mass distribution of the lighter and the heavier 
$\W$ of each events. Figs.~10 and 11 show the Coulomb effects 
on the average light and heavy mass, $\min(m_1,m_2)$ and
$\max(m_1,m_2)$. Below and around threshold almost the whole $\W$ 
mass shift is found in the lighter $\W$: since the lighter $\W$ 
here is off the mass shell, a further change of the mass means
a smaller relative change of the propagator value than a corresponding
change for the heavier $\W$. Mathematically, a $\W$ propagator 
$\rho \simeq 1/((s-\MW^2)^2 + \MW^2\GW^2)$ gives a relative change 
$(1/\rho) |\d \rho / \d s | = 2 |s-\MW^2|/((s-\MW^2)^2 + \MW^2\GW^2)$ 
that is maximal around $\sqrt{s} \simeq \MW \pm \GW/2$ and falls off
like $1/|s-\MW^2|$ in either tail. At larger energies the mass shift
is shared evenly between the two $\W$'s, which here both are  
about equally close to the nominal mass $\MW$.

In the threshold region, there would then be two reasons to base a 
topological determination of the $\W$ mass on the mass distribution
of the heavier $\W$ of the event. Firstly, Coulomb corrections are 
small, and therefore also the associated uncertainty. Secondly,
this mass distribution is significantly narrower that that of the
lighter $\W$ or even that of the average mass 
$(m_{\mrm{heavy}}+m_{\mrm{light}})/2$, Figs. 12, 13 and 14,
so the statistical error would be smaller. However, there are also
two reasons against such a study. Firstly, it would be a major
experimental challenge to reconstruct two $\W$ masses per event,
so the loss in accuracy from experimental effects would almost 
certainly outweight the above gain. Secondly, in this energy region 
a better accuracy can be obtained from a measurement of the total 
cross section \cite{VK1}.

The root-mean-square width of the average $\W$ mass distribution
is narrowest at around 170~GeV, Fig.~14. Below this energy it increses 
significantly for the reasons discussed above; above 170~GeV there is 
a very slight increase from phase space effects. Coulomb effects narrow 
the mass distribution at around threshold, Fig.~15. Again this 
behaviour can be related to the width of the momentum distribution, 
Fig.~16, and to a narrowing of the width when Coulomb effects are 
included, Fig.~17. In general, however, the relation 
between $\sigma_{m} - \sigma_{m,\mrm{nC}}$ and 
$\sigma_{p} - \sigma_{p,\mrm{nC}}$ is not as straightforward
as between $\mavg - \langle m_{\mrm{nC}} \rangle$ and 
$\pavg - \langle p_{\mrm{nC}} \rangle$. This is because the
factor $p$ in the \mbox{$\delta m \approx - p \delta p /m$}  
relation suppresses the importance of the low-momentum tail
on the mass distribution, while it enhances the importance of the 
high-momentum tail. Thus the stable-Coulomb scenario has a broader
momentum distribution than the no-Coulomb one at most energies, 
but since this is due mainly to the enhancement of the low-momentum 
tail, the mass distribution may still be narrower. In the 
unstable-Coulomb approaches, the enhancement is in the high-momentum
tail, and thus the momentum and mass width changes better follow suit.

As we have noted earlier, experimental $\W$ mass methods are based on
the $p$ observable rather than on the $m_1$ and $m_2$ ones. That is,
measured jet/lepton directions and energies are not used to reconstruct
two masses per event. Instead the two $\W$'s are assumed to have the
same mass, and beam energy constraints are used to improve the
detector energy resolution, so that most of the experimental information
is related to angles and from there to the net momentum of jet pairs, 
i.e. to $p$. The translation from a set of $p$ measurements to a $\W$
mass can be done in different ways, that are more or less sensitive to 
Coulomb effects. Compare Figs.~18 and 19; in the former the momentum
$p$ is first averaged over events and then translated to an average mass, 
$\mavg = \sqrt{s/4 - \pavg^2}$, in the latter the $p$'s are first 
translated to $m$'s and then averaged, 
$\mavg = \langle \sqrt{s/4 - p^2} \rangle$. It turns out that the
former recipe is preferrable in the threshold region, in that it
gives a smaller spread between the alternatives, while there is
no noticeable difference at higher energies. 
   
Other averaging schemes can be considered, where also events are giving
relative weights to reflect their ``usefulness'' in constraining the
$\W$ mass. The most direct approach is to remove some fraction of the 
events in the wings of the momentum distributions, where events are
found only if one of the $\W$'s is significantly off the mass shell.
For experimental rejection of background events, such an approach 
offers additional benefits. The importance of the wings for the 
momentum shift of Coulomb effects is shown in Fig.~20 for a few 
energies. What is plotted is 
\begin{equation}
\frac{\d \Delta \pavg}{\d p} = 
\frac{ \d ( \langle p_{\mrm{C}} \rangle - \langle p_{\mrm{nC}}%
\rangle )}{\d p} = (p - \pavg_{\mrm{nC}}) \, 
\left( \frac{1}{\sigma_{\mrm{C}}} \, \frac{\d \sigma_{\mrm{C}}}{\d p} 
- \frac{1}{\sigma_{\mrm{nC}}} \, \frac{\d \sigma_{\mrm{nC}}}{\d p}
\right) ~,
\label{pavgdiff}
\end{equation}
where index ``C'' (``nC'') denotes a scenario with (without) Coulomb
effects. By construction, the measure vanishes at $\pavg_{\mrm{nC}}$;
this is visible at the lower energies but is somewhat hidden by the
finite resolution at the highest energies. In other words, the 
shift of the average momentum is dominated by the behaviour in the 
wings at low energies. At higher energies, the momentum distribution 
is narrower and also the shift of the average momentum is dominated
by a smaller region. This behaviour carries over to the mass shift
$\mavg - \langle m_{\mrm{nC}} \rangle$. We study two alternatives, 
keeping only those events with momentum within 30 and 10~GeV, 
respectively, of the average. For simplicity, $\pavg$ has here been 
approximated by the expression 
$\pavg_{\mrm{approx}} = \sqrt[4]{(s/4-\MW^2)^2 + 4 \MW^2\GW^2}$,
which is not particularly accurate below threshold but good enough
for this purpose. Comparing the mass shifts in Figs.~21 and 22 with 
those in Fig.~9, we note the reduced importance of Coulomb effects 
below and around threshold, while the reduction is rather modest at 
higher energies. For instance, at a typical  LEP~2 energy of 175 GeV, 
the difference between unstable Coulomb and no Coulomb decreases from 
$-23$~MeV without cut to $-16$~MeV with a cut
$| p - \pavg_{\mrm{approx}} | < 10$~GeV. The statistics will drop if 
the momentum window is reduced further, so it appears difficult to 
do much better. 

We have compared several alternative Coulomb effect scenarios in 
this paper. Thereby we hope to learn about the possible effects 
also of other (not yet calculated) QED effects, like 
intermediate--final interference.
Whereas a scenario such as the no-Coulomb one obviously is excluded, 
we have above indicated how something akin to it might arise as
an approximate description of the normalized momentum distribution.  
Such ambiguities would be resolved if the experimental data themselves
could be used to extract the form of QED corrections. One specific 
idea would be to divide the experimental momentum spectrum by the 
no-Coulomb rate, to obtain a distribution akin to what is shown 
in Fig.~4. If the experimental distribution would agree with 
contemporaneous calculations, theory could be used to extract
the $\W$ mass. If not, the experimentally defined correction factors
could still be used for this purpose. 

There are two major problems with such a thinking, however. Firstly, 
so long as we do not know the $\W$ mass we also do not know which 
no-Coulomb curve to compare with. The sharp increase in the 
unstable Coulomb factor at around $p_0$ for the higher energies in 
Fig.~4 is there since we compare scenarios with the same $\MW$. 
If instead $\pavg$ is made to agree, as would be the realistic 
experimental procedure, differences between scenarios are not as
spectacular. Still, even with the same $\pavg$, the shape of the 
momentum spectrum would be different, so with hard work it should
be possible to find which descriptions work and which do not. 
However, here enters the second problem, namely that of limited 
statistics. A LEP~2 experiment will at most have of order 10,000    
events to base the analysis on. It is difficult to see how this 
would be enough to observe any differences, given that we are 
speaking of Coulomb correction factors of a few per cent.

This prejudice is borne out by detailed studies. For instance, in
Fig.~23 we have compared 100 ``LEP~2 experiments'' at 175~GeV, 
each of 10000 events. In the stable-Coulomb scenario the $\MW$ 
parameter has been shifted down by 13~MeV relative to the no-Coulomb
one, and in the unstable-Coulomb one up by 22~MeV (cf.~Fig.~9),
so that $\pavg$ should be the same. In fact, the limited statistics
leads to a non-negligible spread of $\pavg$ values. Also higher
moments, in Fig.~23 $\sqrt[n]{\langle (p - \pavg)^n \rangle}, n=2,3,4$,
show a significant statistical spread, with no obvious separation
between the no-, stable- and unstable-Coulomb scenarios. For a given
``LEP~2 experiments'' it is impossible to tell which distribution it 
is drawn from. Correlations between the moments again do not
distinguish (not shown).

One therefore must conclude that --- unless something truly 
spectacular happens --- experimental input can not be used to pin down
QED interconnection effects. It is necessary to have the theory
under control to the required accuracy. 

\section{Conclusions}

One of the most critical single observables for LEP~2 physics is the 
$\W$ mass. Had there been no interconnection effects between the $\W^+$
and the $\W^-$, in principle their respective four-momenta could be 
reconstructed
and squared to give the $\W^{\pm}$ masses. The shifts in the observable
$\W$ mass distributions could come from a number of sources
\cite{VK6,VK7,Leif} that we have not attempted to address here.
Our main concern has been to estimate the $\W$ mass shift caused by
distortions of the $\W$ momentum spectrum from instability effects,
as embodied in the QED radiative corrections to the differential
$\ee \to \W^+\W^-$ cross section. In the threshold region we can exploit
the formulae for Coulomb corrections \cite{VK2,VK9}, 
which here are responsible for the dominant off-shellness-induced 
phenomena. For complete predictions in the whole LEP~2 energy range,
one needs to incorporate the contributions of other QED interconnection
mechanisms, e.g. those resulting from intermediate--final and
final--final photon interference \cite{VK3,VK8}. 
As far as we are aware, the complete analytical calculation has never
been performed for the whole set of electroweak corrections to unstable
$\W$ production and decay. (An attempt to estimate QED interconnection
phenomena in the relativistic region can be found in ref.~\cite{VK8}.)

While the complete calculation is missing, one may attempt to compare
some different scenarios that give an impression of the size of 
effects. Our current ``best'' description is the all-orders Coulomb
description in ref.~\cite{VK14}. One conclusion of this paper is that 
second-order (and, by implication, all higher-order) Coulomb effects 
are practically negligible. The first-order (``unstable'') Coulomb 
description therefore
is a realistic baseline. Other intermediate--final and final--final 
interference terms are expected to modify this behaviour. Specifically, 
one could expect the rapid variation of the Coulomb factor around $p_0$
(the on-mass-shell $\W$ momentum) to be substantially dampened.
In fact, a brute-force application of a proposed asymptotic dampening
factor \cite{VK8} to the LEP~2 energy region would imply that a
momentum-dependent Coulomb factor can be replaced by an almost
momentum-independent average. Therefore the no-Coulomb scenario  
could be a realistic alternative for reconnection effects on kinematical
distributions. The final scenario, with Coulomb effects evaluated for
stable $\W$'s, is less realistic and has been included mainly for
completeness.  

At LEP~2 energies, the no- and unstable-Coulomb scenarios differ by
about 20~MeV on the observable $\mavg$, given a common theory input $\MW$. 
This is somewhat below the statistical accuracy, but comparable with 
other potential sources of systematic error, and so not negligible. It 
is therefore interesting to see whether the uncertainty can be reduced 
by a clever choice of experimental procedure. The conclusions of this
paper are not too optimistic on this count. The limited statistics
will preclude the testing of the theory scenarios from data itself.
One will therefore be totally dependent on  a full-scale theory 
calculation to reach any definite conclusions. However, different 
statistical treatments of the data can increase or decrease the 20~MeV 
number above by some amount: in the paper we have given the examples of 
cuts on the momentum range analyzed and choice of statistical averaging 
procedure. 

The QED final-state interaction could induce some systematic effects in 
other $\W$-mass measurements as well, for instance in $\pbarp$ collider
experiments. Of particular interest here is the subprocess
$\q\g \to \W \q'$ with $\W \to \ell \nu_{\ell}$. Collider experiments
normally rely on the equivalent process for $\Z^0$ production,
$\q\g \to \Z \q$ with $\Z \to \ell^+ \ell^-$, to calibrate the $\W$
mass scale. Non-universal interference effects are not included in
such a procedure, e.g. a charged ($\W^{\pm}$) versus a chargeless 
($\Z^0$) intermediate state. The effects of real-photon emission
have been compared \cite{Zeppen}, but virtual corrections remain to
be studied. It could well turn out that a corresponding theory 
uncertainty of order 20~MeV exists in this process. Within current
experimental errors this would be negligible, but it could become
relevant for future high-precision measurements.

\subsection*{Acknowledgements}

We are grateful to the UK PPARC for support. Useful discussions with  
V.S.~Fadin, N.~Kjaer, K.~Melnikov, R.~M{\o}ller, W.J.~Stirling and
O.~Yakovlev are acknowledged.
This work was supported in part by the EU Programme
``Human Capital and Mobility'', Network ``Physics at High Energy
Colliders'', contract CHRX-CT93-0319 (DG 12 COMA).

\clearpage

\section*{Figure Captions}
 
\begin{list}{}{\setlength{\leftmargin}{2cm}
  \setlength{\labelwidth}{1.3cm}\setlength{\labelsep}{0.7cm}
  \setlength{\rightmargin}{0cm}}

\item[Fig.~~1]
Total $\ee \to \W^+\W^-$cross section $\sigma$  as a function of the 
c.m. energy. Dashed-dotted curve is no Coulomb, dashed stable Coulomb, 
full unstable Coulomb and dotted second-order Coulomb. 

\item[Fig.~~2]
Total cross section relative to the no-Coulomb one, $\sigma/\sigma_0$.
Notation as in Fig.~1.

\item[Fig.~~3] 
The differential cross section $\d \sigma / \d p$ in pb/GeV. 
Notation as in Fig.~1. 

\item[Fig.~~4]
Ratio of the differential cross sections with and without Coulomb 
effects included, 
$(\d \sigma / \d p)_{\C} / (\d \sigma / \d p)_{\mathrm{nC}}$.
Dashed curve is stable Coulomb, full unstable Coulomb and dotted 
second-order Coulomb. At the two highest energies, dashed-dotted
curve is dampened Coulomb.

\item[Fig.~~5]
Ratio of second- to first-order Coulomb corrections,
$(|f(\mathbf{p},E)|^2 - 1 - \alpha\overline{\delta}_{\C}/\beta) / 
(\alpha\overline{\delta}_{\C}/\beta)$,
as a function of $p$. The dashed line at small $p$ indicates the
$p \to 0$ limit given in eq.~(\ref{limitsmallp}).

\item[Fig.~~6]
Average $\W$ momentum, $\pavg$. 
Notation as in Fig.~1.

\item[Fig.~~7]
Shift of the average $\W$ momentum relative to the no-Coulomb value,
\mbox{$\pavg - \langle p_{\mrm{nC}} \rangle$}.
Notation as in Fig.~1.

\item[Fig.~~8]
Shift of the average $\W$ mass relative to the nominal one,
$\mavg - \MW$. Notation as in Fig.~1.

\item[Fig.~~9]
Shift of the average $\W$ mass relative to the no-Coulomb value,
$\mavg - \langle m_{\mrm{nC}} \rangle$.
Notation as in Fig.~1. 

\item[Fig.~10]
Shift of the lighter $\W$ mass relative to the no-Coulomb value,
$(\mavg - \langle m_{\mrm{nC}} \rangle)_{\mrm{light}}$.
Notation as in Fig.~1.

\item[Fig.~11]
Shift of the heavier $\W$ mass relative to the no-Coulomb value,
\mbox{$(\mavg - \langle m_{\mrm{nC}} \rangle)_{\mrm{heavy}}$}.
Notation as in Fig.~1.

\item[Fig.~12]
Root-mean-square width of the lighter $\W$ mass distribution, 
$\sigma_{m,\mrm{light}}$.
Notation as in Fig.~1.

\item[Fig.~13]
Root-mean-square width of the heavier $\W$ mass distribution, 
$\sigma_{m,\mrm{heavy}}$.
Notation as in Fig.~1.

\item[Fig.~14]
Root-mean-square width of the average $\W$ mass distribution, 
$\sigma_{m}$.
Notation as in Fig.~1.

\item[Fig.~15]
Shift of the width of the average $\W$ mass distribution relative to 
the no-Coulomb value $\sigma_{m} - \sigma_{m,\mrm{nC}}$.
Notation as in Fig.~1.

\item[Fig.~16]
Root-mean-square width of the $\W$ momentum distribution, $\sigma_p$.
Notation as in Fig.~1.

\item[Fig.~17]
Shift of the width of the $\W$ momentum distribution relative to the 
no-Coulomb value, $\sigma_p - \sigma_{p,\mrm{nC}}$. 
Notation as in Fig.~1.

\item[Fig.~18]
Shift of the average $\W$ mass relative to the no-Coulomb value,
$\mavg - \langle m_{\mrm{nC}} \rangle$. Here the 
average momentum $\pavg$ has been converted into an average mass
$\mavg = \sqrt{s/4 - \pavg^2}$.  
Notation as in Fig.~1.

\item[Fig.~19]
Shift of the average $\W$ mass relative to the no-Coulomb value,
$\mavg - \langle m_{\mrm{nC}} \rangle$. Here the 
momentum $p$ in each event has been converted into an effective mass
$m = \sqrt{s/4 - p^2}$, and this mass has been averaged over events.  
Notation as in Fig.~1.

\item[Fig.~20]
The buildup of the total shift in $\pavg$ as a function
of $p$, as given in eq.~(\ref{pavgdiff}). The result is given in 
relation to the no-Coulomb alternative. Dashed curve is stable Coulomb, 
full unstable Coulomb and dotted second-order Coulomb.

\item[Fig.~21]
Shift of the average $\W$ mass relative to the no-Coulomb value,
$\mavg - \langle m_{\mrm{nC}} \rangle$. Here only 
events with $| p - \pavg_{\mrm{approx}} | < 30$~GeV have been included.
Notation as in Fig.~1.

\item[Fig.~22]
Shift of the average $\W$ mass relative to the no-Coulomb value,
$\mavg - \langle m_{\mrm{nC}} \rangle$. Here only 
events with $| p - \pavg_{\mrm{approx}} | < 10$~GeV have been included.
Notation as in Fig.~1.

\item[Fig.~23]
The distribution of some moments of the momentum distribution in
100 ``LEP~2 experiments'' of 10000 events each at 175~GeV. 
Dashed-dotted histogram is no Coulomb, dashed stable Coulomb, 
and full unstable Coulomb. The $\MW$ mass parameter has been adjusted
to give the same $\pavg$ in the three scenarios.   

\end{list}

\clearpage
\begin{center}
\mbox{\epsfig{file=lutp95181.epspm}}\\
Figure 1
\end{center}  
\begin{center}
\mbox{\epsfig{file=lutp95182.epspm}}\\
Figure 2
\end{center}  

\clearpage
\begin{center}
\mbox{\epsfig{file=lutp95183.epspm}}
\\[0.5cm]
Figure 3
\end{center}

\clearpage
\begin{center}
\mbox{\epsfig{file=lutp95184.epspm}}
\\[0.5cm]
Figure 4
\end{center}

\clearpage
\begin{center}
\mbox{\epsfig{file=lutp95185.epspm}}
\\[0.5cm]
Figure 5
\end{center}

\clearpage
\begin{center}
\mbox{\epsfig{file=lutp95186.epspm}}\\
Figure 6
\end{center}  
\begin{center}
\mbox{\epsfig{file=lutp95187.epspm}}\\
Figure 7
\end{center}  

\clearpage
\begin{center}
\mbox{\epsfig{file=lutp95188.epspm}}\\
Figure 8
\end{center}  
\begin{center}
\mbox{\epsfig{file=lutp95189.epspm}}\\
Figure 9
\end{center}  

\clearpage
\begin{center}
\mbox{\epsfig{file=lutp951810.epspm}}\\
Figure 10
\end{center}  
\begin{center}
\mbox{\epsfig{file=lutp951811.epspm}}\\
Figure 11
\end{center}  

\clearpage
\begin{center}
\mbox{\epsfig{file=lutp951812.epspm}}\\
Figure 12
\end{center}  
\begin{center}
\mbox{\epsfig{file=lutp951813.epspm}}\\
Figure 13
\end{center}  

\clearpage
\begin{center}
\mbox{\epsfig{file=lutp951814.epspm}}\\
Figure 14
\end{center}  
\begin{center}
\mbox{\epsfig{file=lutp951815.epspm}}\\
Figure 15
\end{center}  

\clearpage
\begin{center}
\mbox{\epsfig{file=lutp951816.epspm}}\\
Figure 16
\end{center}  
\begin{center}
\mbox{\epsfig{file=lutp951817.epspm}}\\
Figure 17
\end{center}  

\clearpage
\begin{center}
\mbox{\epsfig{file=lutp951818.epspm}}\\
Figure 18
\end{center}  
\begin{center}
\mbox{\epsfig{file=lutp951819.epspm}}\\
Figure 19
\end{center}  

\clearpage
\begin{center}
\mbox{\epsfig{file=lutp951820.epspm}}
\\[0.5cm]
Figure 20
\end{center}  

\clearpage
\begin{center}
\mbox{\epsfig{file=lutp951821.epspm}}\\
Figure 21
\end{center}  
\begin{center}
\mbox{\epsfig{file=lutp951822.epspm}}\\
Figure 22
\end{center}  

\clearpage
\begin{center}
\mbox{\epsfig{file=lutp951823.epspm}}
\\[0.5cm]
Figure 23
\end{center}  
\begin{picture}(0,0)(0,0)
\put(100,380){$\pavg$ [GeV]}
\put(300,380){$\sqrt{\langle (p - \pavg)^2 \rangle}$ [GeV]}
\put(80,67){$\sqrt[3]{\langle (p - \pavg)^3 \rangle}$ [GeV]}
\put(300,67){$\sqrt[4]{\langle (p - \pavg)^4 \rangle}$ [GeV]}
\end{picture}

\end{document}